\documentclass{article}

\usepackage{arxiv}

\usepackage[utf8]{inputenc} 
\usepackage[T1]{fontenc}    
\usepackage{hyperref}       
\usepackage{url}            
\usepackage{booktabs}       
\usepackage{amsfonts}       
\usepackage{nicefrac}       
\usepackage{microtype}      
\usepackage{lipsum}
\usepackage{graphicx}
\graphicspath{ {./images/} }

\title{Three-Dimensional Single-Shot Ptychography}

\author{
 David Goldberger \\
  Department of Physics\\
  Colorado School of Mines\\
  Golden, CO 80401 \\
  \texttt{dgoldberger@mines.edu} \\
   \And
 Jonathan Barolak \\
  Department of Physics\\
  Colorado School of Mines\\
  Golden, CO 80401 \\
  \And
 Charles G. Durfee \\
  Department of Physics\\
  Colorado School of Mines\\
  Golden, CO 80401 \\
   \And
 Daniel E. Adams \\
  Department of Physics\\
  Colorado School of Mines\\
  Golden, CO 80401 \\
}

\begin{document}
\maketitle
\begin{abstract}
Here we introduce three-dimensional single-shot ptychography (3DSSP). 3DSSP leverages an additional constraint unique to the single-shot geometry to deconvolve multiple 2D planes of a 3D object. Numeric simulations and analytic calculations demonstrate that 3DSSP reconstructs multiple planes in an extended 3D object with a minimum separation consistent with the depth of field for a conventional microscope. We experimentally demonstrate 3DSSP by reconstructing orthogonal hair strands axially separated by 5 mm. Three-dimensional single-shot ptychography provides a pathway towards volumetric imaging of dynamically evolving systems on ultrafast timescales. 
\end{abstract}


\section{Introduction}
\begin{figure}[htb]
\centering
\includegraphics[width=\linewidth]{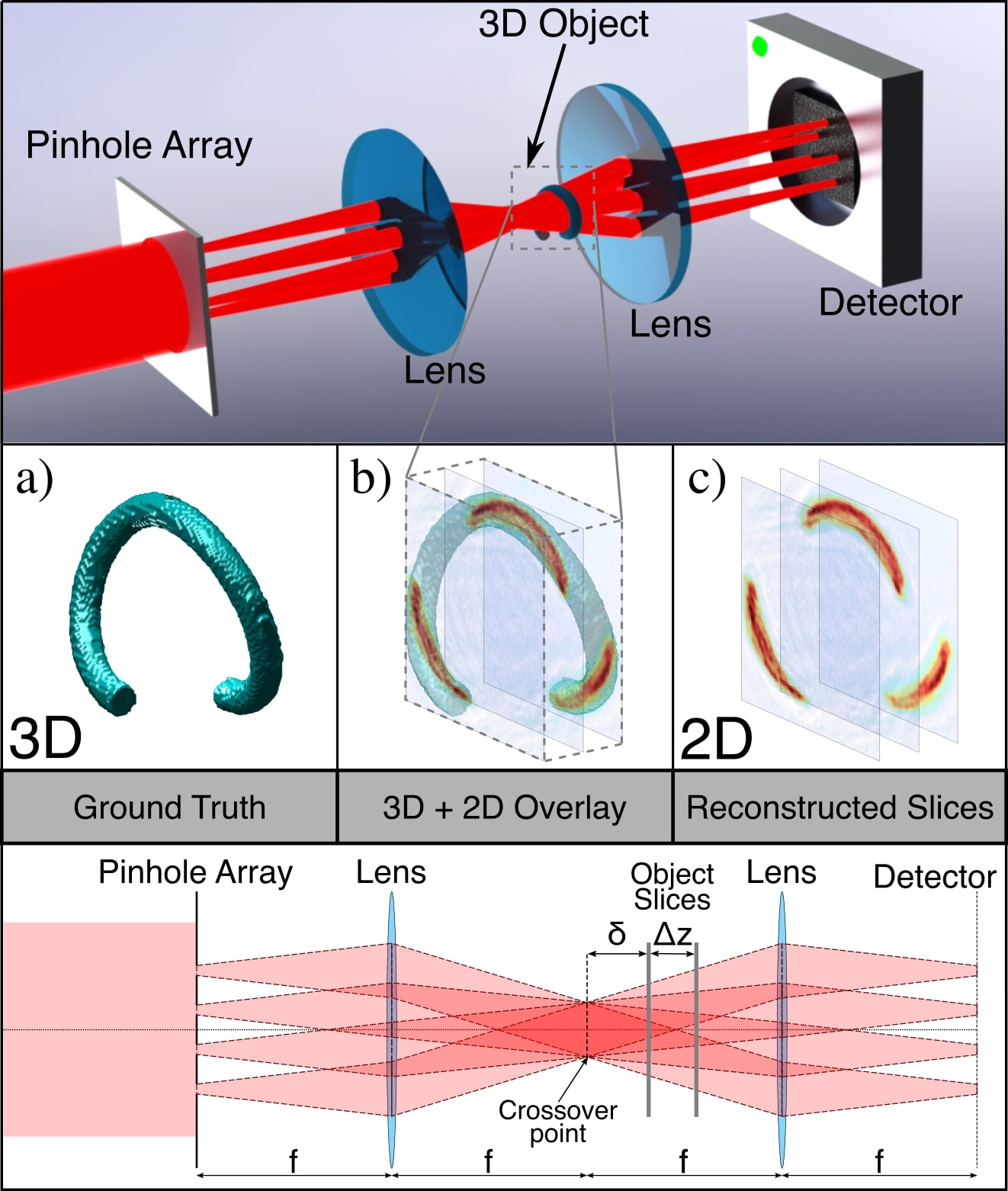}
\caption{Schematic representation of the 3DSSP experiment and results from preliminary simulations. The top panel shows a diagram of the experimental system, while the bottom panel shows a ray tracing schematic. The experiment consists of incoming plane-wave illumination, a DOE, 4F imaging system and detector. The object is placed a distance $\delta$ away from the crossover point, and the inter-plane spacing is $\Delta z$. The detector is segmented to record individual diffraction patterns, which are fed into the 3DSSP algorithm along with the initial positions of the beamlets on the first object slice. The middle panel of the figure shows a simulated example reconstruction of three planes of a continuous 3D object. The reconstructed object planes match up well with the continuous 3D object. Interestingly, the broken loop object that is simulated here is not recoverable by standard 2D projection-based imaging techniques.}
\label{fig:setupAndSim}
\end{figure}

Coherent Diffractive Imaging (CDI) is a computational microscopy technique that uses a sophisticated algorithm to processes far-field intensity measurements simultaneously returning both the phase and amplitude of the probing illumination and the specimen \cite{Fienup1982a,Fienup1978a,Elser2003}. The technique is particularly exciting because it can provide simultaneous amplitude and phase contrast imaging across the EM spectrum in both reflection and transmission modalities \cite{Seaberg2011,Clark2011,Porter2017}. In ptychography, an advanced implementation of CDI, a sample is scanned transverse to a probe illumination producing diffraction patterns from overlapping regions of illumination. Transverse scanning of a probe illumination limits the minimum collection time but increases the technique’s robustness and removes the need for \textit{a priori} knowledge of the specimen \cite{Thibault2009,Maiden2009}. Recently developed single-shot ptychography (SSP) eliminates the need for scanning by introducing a diffractive optical element (DOE), often an array of pinholes, and a 4f imaging system  \cite{Pan2013,Sidorenko2016a}. A ray tracing schematic of a conventional SSP imaging system is shown in the bottom panel of Fig. \ref{fig:setupAndSim}. The DOE breaks up incoming illumination into several beamlets. The 4f system collimates the beamlets and then crosses them. The object is slightly offset from the crossover point (which is also the Fourier plane of the 4f imaging system) and the detector is segmented such that the diffraction pattern produced by each beamlet is essentially recorded independently. The system thus provides the same set of diffraction patterns as scanning ptychography but in a single shot. SSP dramatically reduces the acquisition time of a ptychographic dataset and heralds the possibility of time-resolved, simultaneous phase and amplitude contrast imaging, as recently  demonstrated experimentally \cite{Wengrowicz2019,Sidorenko2017}. 

Generally, ptychography depends on the projection approximation, which models the interaction of an illuminating probe with a specimen’s complex transmission function as a product. This approximation is accurate in the limit of an optically thin object \cite{Thibault2008}, but it restricts the technique to imaging effectively two-dimensional (2D) samples. Scanning ptychography has been adapted to image in three-dimensions (3D) using a multi-slice approach (3PIE) and ptychographic tomography  \cite{Li2018,Godden2014,Maiden2012}. These techniques were applied to image thick samples like biological tissues with relatively simple tabletop experimental systems.  While these methods are powerful and robust, their required acquisition times preclude them from measuring dynamic objects. As we probe more complex phenomena at ever shorter time scales the need for metrologies capable of imaging dynamically evolving objects becomes critical. Here we introduce Three-dimensional Single-Shot Ptychography (3DSSP), which uses the same SSP experimental configuration but implements a novel algorithm that leverages an additional object domain constraint to reconstruct multiple 2D planes of a 3D object. Numerical simulations show that 3DSSP can reconstruct multiple planes in an extended 3D object with minimum axial separation consistent with the depth of field for a conventional microscope. We experimentally demonstrate 3DSSP by reconstructing orthogonal hair strands axially separated by 5 mm. This  novel imaging technique provides a pathway toward volumetric imaging of dynamically evolving systems on ultrafast timescales, such as the nonlinear response of materials \cite{Odstrcil2016,Bernert2017,Adams2010a}.

\section{Methods}
\label{sec:headings}
Current single-shot ptychographic experiments depend on reconstruction algorithms developed for conventional ptychography like ePIE  \cite{Maiden2009,Sidorenko2016a}. Conventional reconstruction algorithms provide sufficient results when applied to 2D single-shot experiments as long as the object is optically thin \cite{Thibault2008}, but thick objects require more sophisticated multi-slice reconstruction algorithms. However, the geometry of the single-shot system prohibits direct application of conventional multi-slice algorithms like 3PIE \cite{Maiden2012}. In particular, the beamlets are collimated and propagate at an angle through the object. Our 3DSSP technique reconstructs 3D objects based on recorded diffraction patterns from a SSP experiment by taking into account the SSP geometry and applying a multi-slice approach. 3PIE was inspired by electron microscopy techniques  \cite{Maiden2012}, and it has been successfully applied to reconstruct phase and amplitude contrast images of continuous 3D objects  \cite{Godden2014}. Like 3PIE, 3DSSP computes exit waves for each 2D slice (or z-slice) of a 3D object, but where 3PIE propagates the exit wave from one slice to form the illumination function at the next slice, 3DSSP propagates $\textit{and shifts}$ each exit wave to account for the geometry of a SSP setup.

\subsection{Inter-slice Propagation}
In SSP an object is placed a distance $\delta$ away from the crossover point. Beamlets created by a DOE are collimated, crossed, and then separate geometrically, as shown in Fig. \ref{fig:setupAndSim}. A critical input to any ptychographic reconstruction algorithm is the position of the probes on the object. In the single-shot geometry, these positions change as a function of $\delta$. To account for transverse shifts between z-slices, we define the following Inter-Slice Propagator (\textit{ISP}) as:
\begin{equation}
\psi_{i_{s+1}}(r) = ISP \{\psi_{e_s}(r) \} \equiv \mathfrak{F}^{-1}\{\mathfrak{F}\{\psi_{e_s}(r)\} \mathcal{H}(f)\}.
\label{eq:ISP}
\end{equation}

\noindent Here $\mathfrak{F}\{\cdot\}$ is the two-dimensional discrete Fourier transform, $\psi_{e_s}$ is the exit surface wave at slice $s = \{1,...,N_s\}$, $N_s$ is the total number of slices, $\psi_{i_{s+1}}$ is the incident surface wave at slice $s+1$, $\vec{f} \equiv (f_x, f_y)$ are spatial frequency grids for the object domain grids $\vec{r} \equiv (x, y)$, $k_0$ is the vacuum angular wave number, and the transfer function $\mathcal{H}$ is given by:
\begin{equation}
\mathcal{H}(f_x,f_y) \equiv e^{i k_0 \Delta z \sqrt{1-\frac{4 \pi^2}{k_0^2}(f_x^2 + f_y^2)}} e^{-2 \pi i (f_x \Delta x + f_y \Delta y)}.
\label{eq:ISPTF}
\end{equation}
\noindent The transfer function in Eqn. \ref{eq:ISPTF} is simply a product of the free-space transfer function \cite{goodman2005introduction} with a linear phase that accounts for the shift between planes including sub-pixel contributions. The magnitudes of the shifts are give by, $\Delta x = \Delta z X/f$ and $\Delta y = \Delta z Y/f$. Where $X$ and $Y$ are the position of a beamlet on the detector, $\Delta z$ is the propagation distance between z-slices for a specific beamlet, which in the small angle approximation is the same for all beamlets, and $f$ is the focal length of the second lens. The positions of beamlets at the first object slice are determined from an image of the DOE, $\delta$, and the focal length of the second lens \cite{Sidorenko2016a}. The inverse inter-slice propagator is then defined as:
\begin{equation}
\psi_{e_{s}}({r}) = ISP^{-1} \{\psi_{i_{s+1}}({r}) \} \equiv \mathfrak{F}^{-1}\{\mathfrak{F}\{\psi_{i_{s+1}}({r})\} \mathcal{H}^* ({f})\}.
\label{eq:InvISP}
\end{equation}
\noindent where $*$ is the complex conjugate. Accounting for the position change of the beamlets at each object slice enables 3D reconstructions via a multi-slice approach for the single-shot geometry. In comparison to 3PIE, 3DSSP places an additional constraint on the problem because the positions at each slice shift by a known amount. However, because the beamlets separate from each other in the SSP geometry, the total 3D volume that can be imaged by 3DSSP is limited.

\subsection{The 3DSSP Algorithm}
\begin{figure}[htb]
\centering
\includegraphics[width=\linewidth]{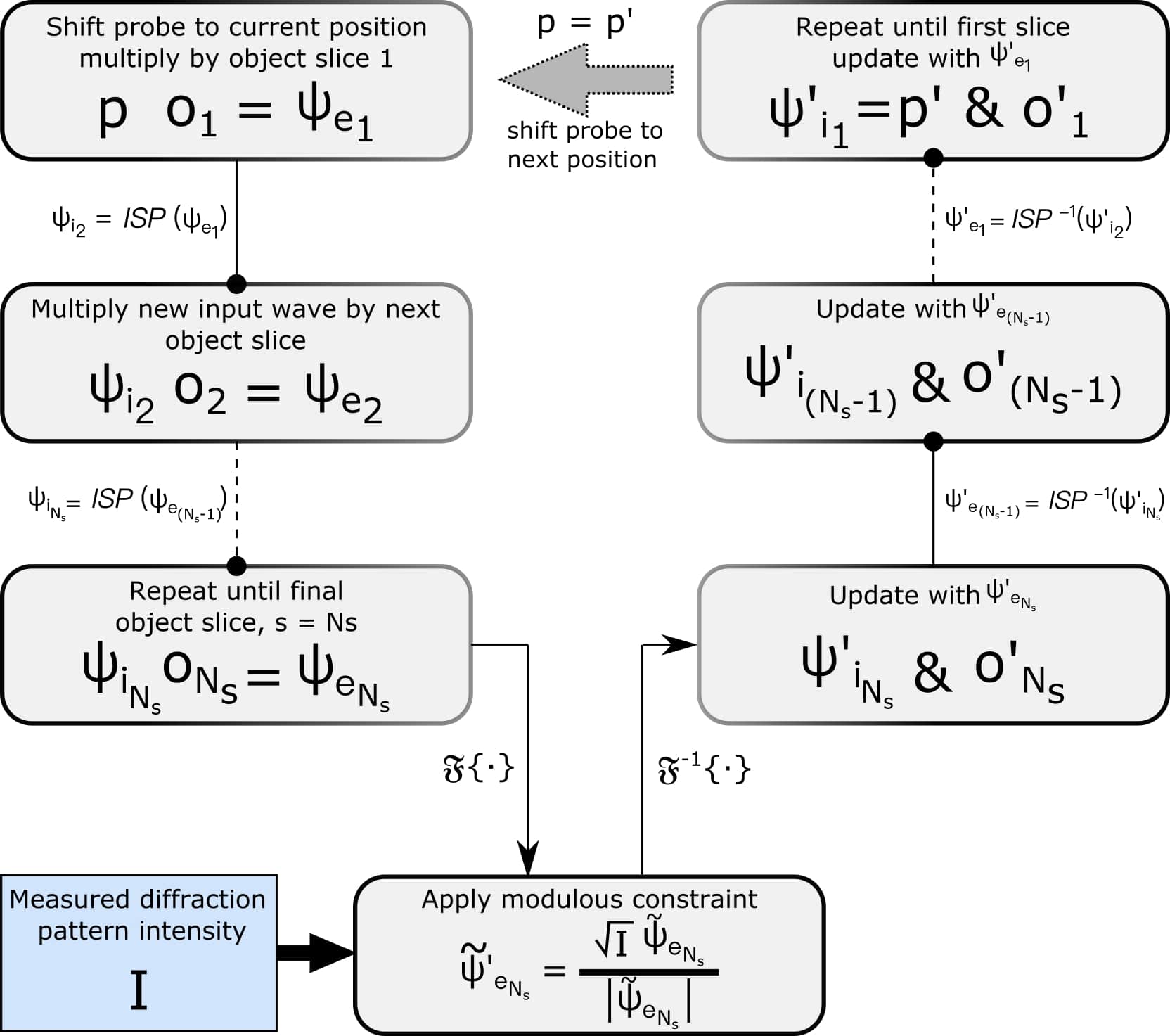}
\caption{Flow diagram of the 3DSSP algorithm. Solids lines terminating in circles below represent application of the Inter-Slice Propagator (\textit{ISP}), while those terminating in circles above represent application of the inverse Inter-Slice Propagator (\textit{ISP}\textsuperscript{-1}). Dashed lines represent repeated application of the previous steps. The downward facing arrow represents the two-dimensional discrete Fourier transform, and the upward facing arrow represents its inverse. Selected steps of the algorithm are shown in the grey curved rectangles and the measured diffraction pattern intensity input is shown in the blue rectangle. After the steps shown are repeated for each beamlet position, the algorithm is said to have completed one full iteration.  }
\label{fig:flowChart}
\end{figure}

In this section, we describe one iteration of the algorithm for a single beamlet and highlight deviations from other imaging methods that make three-dimensional imaging possible in the single-shot geometry. A flow diagram of the 3DSSP algorithm is shown Fig. \ref{fig:flowChart} for reference. For a full reconstruction, the steps presented below are repeated for each beamlet and then multiple overall iterations until some prescribed number of total iterations are completed or until the error reduces below some threshold \cite{Adams2012a, Maiden:17}. The 3DSSP algorithm forms the exit wave at the first z-slice as the product between the incident probe and first slice of the object as $\psi_{e_1} = p \, o_1$. We initialize the algorithm with a guess for the probe by taking the average of the inverse Fourier transforms of all of the segmented DOE beamlet images without the object present. The object is initialized as free space at each slice. The exit waves at subsequent z-slices are then calculated using:
\begin{equation}
\psi_{e_s}(x,y) = \psi_{i_s}(x,y)o_s(x,y),
\label{eq:psi_e_1}
\end{equation}
\noindent with $\psi_{i_1} \equiv p$ is the input probe illumination. This exit wave is then propagated and shifted to form the incident wave for the next slice as:
\begin{equation}
\psi_{i_{s+1}}(x,y) = ISP\{ \psi_{e_s}(x,y) \}.
\label{eq:psi_i_2}
\end{equation}
Eqns. \ref{eq:psi_e_1} and \ref{eq:psi_i_2} are applied repetitively until the exit wave for the last slice is formed i.e., $s = N_s$. The exit wave at the last slice is updated with the measured data according to:
\begin{equation}
\psi_{e_{N_s}}^\prime = \mathfrak{F}^{-1}\{\sqrt{I} \frac{\tilde{\psi}_{e_{N_s}}}{|\tilde{\psi}_{e_{N_s}}|}\},
\label{eq:modConst}
\end{equation}
\noindent where $\psi_{e_{N_s}}^\prime$ is the updated exit wave at slice $N_s$, $\tilde{\psi}_{e_{N_s}} = \mathfrak{F}\{\psi_{e_{N_s}}\}$ and $\textit{I}$ is the measured diffraction pattern intensity from the section of the segmented detector that corresponds to the current beamlet. The modulus constraint in Eqn. \ref{eq:modConst} is a common feature of all CDI algorithms however, in 3DSSP the measured intensity for a given position is a segment of the full detector. The exit waves are back propagated through the object using the inverse ISP in Eqn. \ref{eq:InvISP} and the object and incident waves are updated using the corresponding exit surface wave according to Eqn. 3 from \cite{Maiden2012}.

\subsection{Axial Resolution and Reconstruction Volume}
An important parameter in any 3D imaging system is the axial resolution. The 3DSSP algorithm separates planes in the object because the position of the beamlets change at each slice. Thus, the minimum resolvable separation of planes should be achieved when the position of the outer most beamlet shifts by one transverse resolution unit between two planes separated by the axial resolution, $\delta z$. Based on this premise we calculate the expected axial resolution as:
\begin{equation}
\delta z = \frac{2 \lambda_0 f^2}{X_{det} X_{dif}},
\label{eq:axialRes}
\end{equation}
\noindent where $\lambda_0$ is the vacuum wavelength, $f$ is the focal length of the second lens in the 4f imaging system, and $X_{det}$, $X_{dif}$ are  the size of the full detector and the size of one recorded diffraction pattern, respectively. Recall that these sizes are not the same in the single-shot geometry because the detector is segmented to record diffraction patterns independently. Using the parameters from our experimental single-shot system: $f$ = 5 cm, $\lambda_0$ = 532 nm, $X_{det}$ = 10.85 mm, $X_{dif}$ = 1.59 mm, we calculate an axial resolution of $\delta z$ $\sim$ 150 $\mu$m. It is worth noting that if $X_{det} = X_{dif}$ Eqn. \ref{eq:axialRes} gives the paraxial depth of field for a microscope with magnification much greater than one.

We now calculate a value for the imaging volume in 3DSSP by first noting that the overlap fraction between beamlets, $\eta \equiv A/\pi r^2$, changes as a function of the distance from the crossover point, $z$. Here, $A$ is the overlap area shared between two beamlets and $r$ is the radius of each beamlet. It is common in ptychography to use overlap fractions between $\eta =$ 0.9 and $\eta =$ 0.6, which we use to set the minimum and maximum $z$ values, respectively. Our task is to write an expression, $z(\eta)$, that provides bounds on the axial extent of the imaging volume.

The specific relationship for $z(\eta)$ depends on the shape of DOE elements and their relative arrangement. Here we derive the relation for a DOE similar to our experiment which is an array of circular pinholes of radius, $r_p \sim 32 \mu$m, arranged in a Fermat spiral pattern \cite{Huang2014}. Beamlets passing between the two lenses in the single-shot configuration have a cross-section given by the 2D Fourier transform of a DOE element. This means that the shape of each beamlet when passing between the 4f lenses is the Airy pattern \cite{Airy1835}, which has a central Airy disk that we take to be the illumination function. The Airy disk is surrounded by a circular boundary where the field amplitude drops to zero, which defines the beam radius. The area of overlap between two Airy disks may be approximated for small center-to-center distances, $d$ as: $A \approx \pi r^2 - 2 r d$. In this approximation, the value for the area becomes more accurate with increasing overlap and reaches 10$\%$ error when $\eta \sim 0.35$. The Vogel model for the Fermat spiral  gives the radial distance to each pinhole as $ X_{det}/2 \times \sqrt{n/N_p}$ \cite{Vogel1979}, where $n \in \{0,...,N_p-1\}$ is the index number of each pinhole and $N_p$ is the total number of pinholes. The angular position of each pinhole is given by $ n \, \theta_g$, where $\sqrt{2 \pi/\theta_g} = \varphi$, is the golden ratio.

We approximate the distance between nearest neighbor pinholes as the distance between the $n=0$ and $n=1$ pinhole. This approximation is only valid near the center of the Fermat pattern but this is the region where the reconstruction will have the highest fidelity in any case. The center-to-center distance between Airy disks as they propagate from the crossover point to the second lens is then given by: $d = X_{det}/(2\sqrt{N_p}) \times z/f$.

Writing the approximate overlap area as $A = \eta \pi r^2$, substituting $d$, and solving for $z$ gives:

\begin{equation}
z(\eta) = \pi \, 0.61 \, \delta z \, \frac{X_{dif}}{2 r_p} \, \sqrt{N_p} \, (1-\eta),
\label{eq:axialDist}
\end{equation}

\noindent where we have used $r = 0.61 \lambda_0 f/r_p$ to relate the radius of the Airy disk to the pinhole radius. Using Eqn. \ref{eq:axialDist}, we calculate an axial extent, $z(\eta = 0.6) - z(\eta = 0.9) \sim $ 1.4 cm. This extent sets the height of a truncated cone that fills a volume of $\sim$ 75 mm$^3$.

\section{Results}
\subsection{Simulated Experiments}
In the simulations presented here we calculate ideal diffraction patterns using parameters similar to our experimental system and the forward model in Eqns. \ref{eq:psi_e_1} and \ref{eq:psi_i_2}. The simulated 4f system has unit magnification with $f$ = 5 cm focal lengths for each lens. The detector has 2048$\times$2048 square pixels with a $dX$ = 5.3 $\mu$m pitch and the input wavelength was $\lambda_0$ = 532 nm. This produces a collimated beam diameter between the lenses given by $D \sim$ 0.5 mm. With these parameters the oversampling is calculated as $\sigma = f \lambda_0/D dX$, which corresponds to a value of 5. In addition to satisfying oversampling, 3DSSP requires the distance from the crossover point to each slice in the reconstruction. In the simulated and experimental results presented here, we specify the distance to each object slice as the distance from the crossover point to the first slice, $\delta$, plus the distance between slices, $\Delta z$.

\begin{figure}[htbp]
\centering
\includegraphics[width=\linewidth]{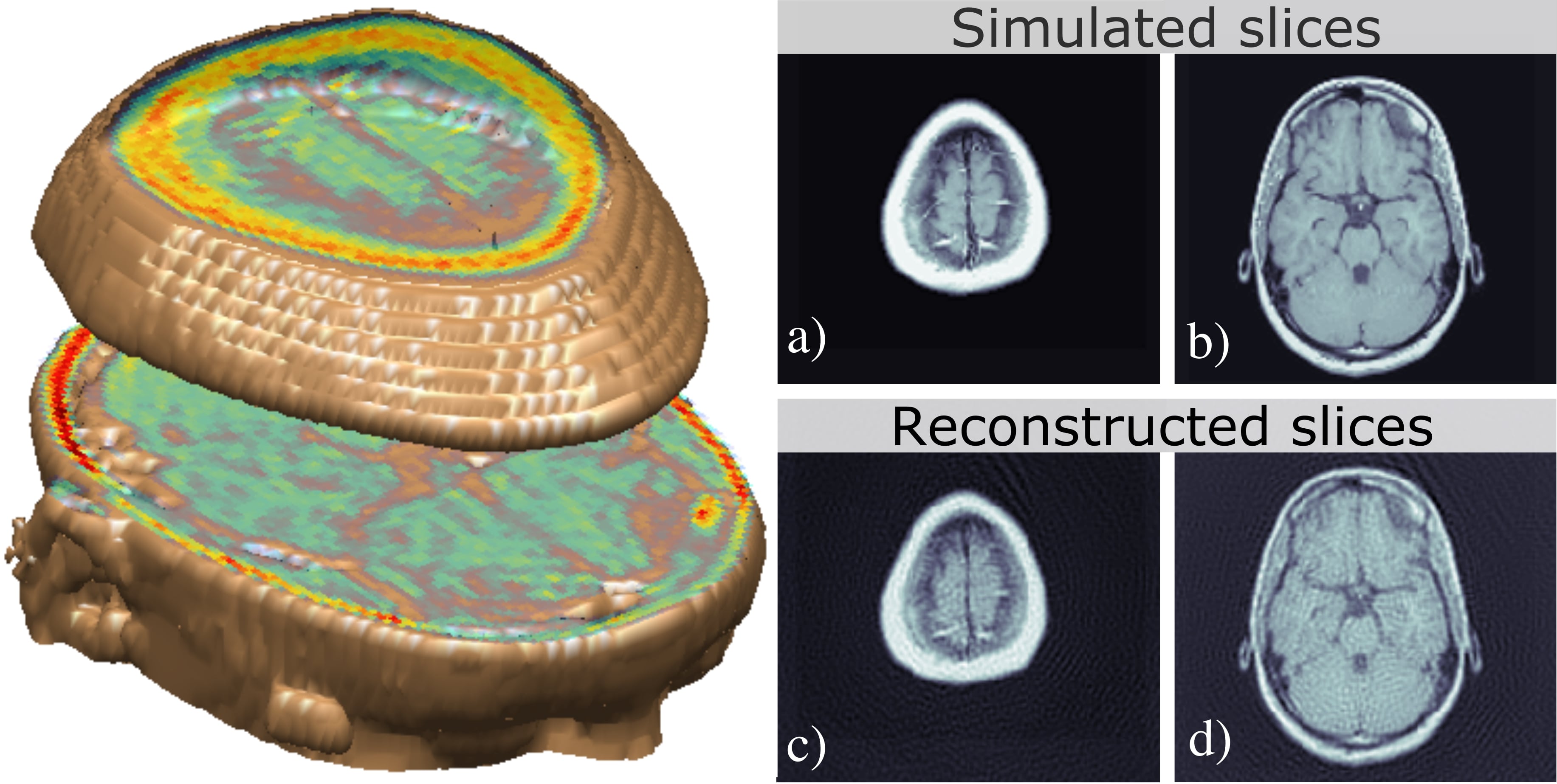}
\caption{Simulated 3DSSP experiment. a) and b) show the simulated slices used to produce ideal diffraction patterns while c) and d) are the corresponding reconstructions. Note that the fine details of the simulated slices are well represented by the reconstructed slices. The left panel shows the reconstructed slices superimposed on the head geometry. The reconstructed slices fit as expected showing that the reconstructed slices also accurately account for the changing size of each slices. This promising result suggests that the 3DSSP technique can be applied to continuously varying 3D objects with fine detail and good reconstruction quality.}
\label{fig:mriSim}
\end{figure}

The object was scaled to fit within the theoretical field of view at each plane. Then we simulated the input illumination using parameters of our experimental DOE and 4f imaging system. For the simulations presented here, the DOE is a Fermat spiral of $N_p =$ 40 identical pinholes. A Fermat spiral DOE lends itself nicely to SSP experiments because it gives the most uniformly overlapped probes on the object \cite{Huang2014}. We Fourier transform the DOE to calculate the total illumination at the crossover point and then use the free space transfer function to propagate the illumination a distance $\delta$ to the first object slice.

We model the interaction of this illumination with the first 2D slice of the object by multiplying it by the complex transfer function of that slice, and define the result as the first exit wave. Then we propagate the first exit wave to the second slice and multiply it by the complex transfer function of that slice. We repeat this process $N_s$ times to calculate the final exit wave. A two-dimensional discrete Fourier transform of the final exit wave gives a simulated diffraction pattern. Finally, the detector is segmented into $N_p$ sections by applying a centroidal Voronoi tessellation, which gives the maximum area for each beamlet's diffraction pattern. Together the $N_p$ simulated diffraction patterns make a ptychographic data set, which we feed into the 3DSSP reconstruction algorithm.

Fig. \ref{fig:setupAndSim} presents the results from our first simulated experiment where the 3D object is a continuous broken loop structure. We simulated and reconstructed three slices taken from the continuous 3D object each separated by 3 mm. The 3DSSP algorithm successfully reproduced the slices of the object. This object is particularly interesting because it is inherently anisotropic and standard 2D projection-based imaging techniques like Schlieren imaging or interferometric imaging would have incorrectly produced an image of a ring. Attempts to use Abel inversion to produce a 3D image from 2D projections would also fail.

The first simulation shows the power of 3DSSP as the algorithm successfully reconstructs the object where other techniques fail. However, the object in that simulation is relatively simple as it does not contain fine details and its size does not change significantly. To investigate the quality of reconstructions from the 3DSSP algorithm, we performed a second simulated experiment in which the object was two slices from a MATLAB example MRI scan separated by 1.5 mm. We chose the MRI scan slices as an object because they contain small features and the size of the images continuously vary. The two slices that we used to simulate diffraction patterns are shown in a) and b) of Fig. \ref{fig:mriSim} above, while the reconstructed slices are shown in c) and d). The left panel of Fig. \ref{fig:mriSim} shows the reconstructed slices overlaid on the head. While the scale of a human head is much larger than the field of view of our current imaging system, this simulation shows that the 3DSSP algorithm is capable of capturing fine details from continuously varying 3D samples.

\begin{figure}[htb]
\centering
\includegraphics[width=\linewidth]{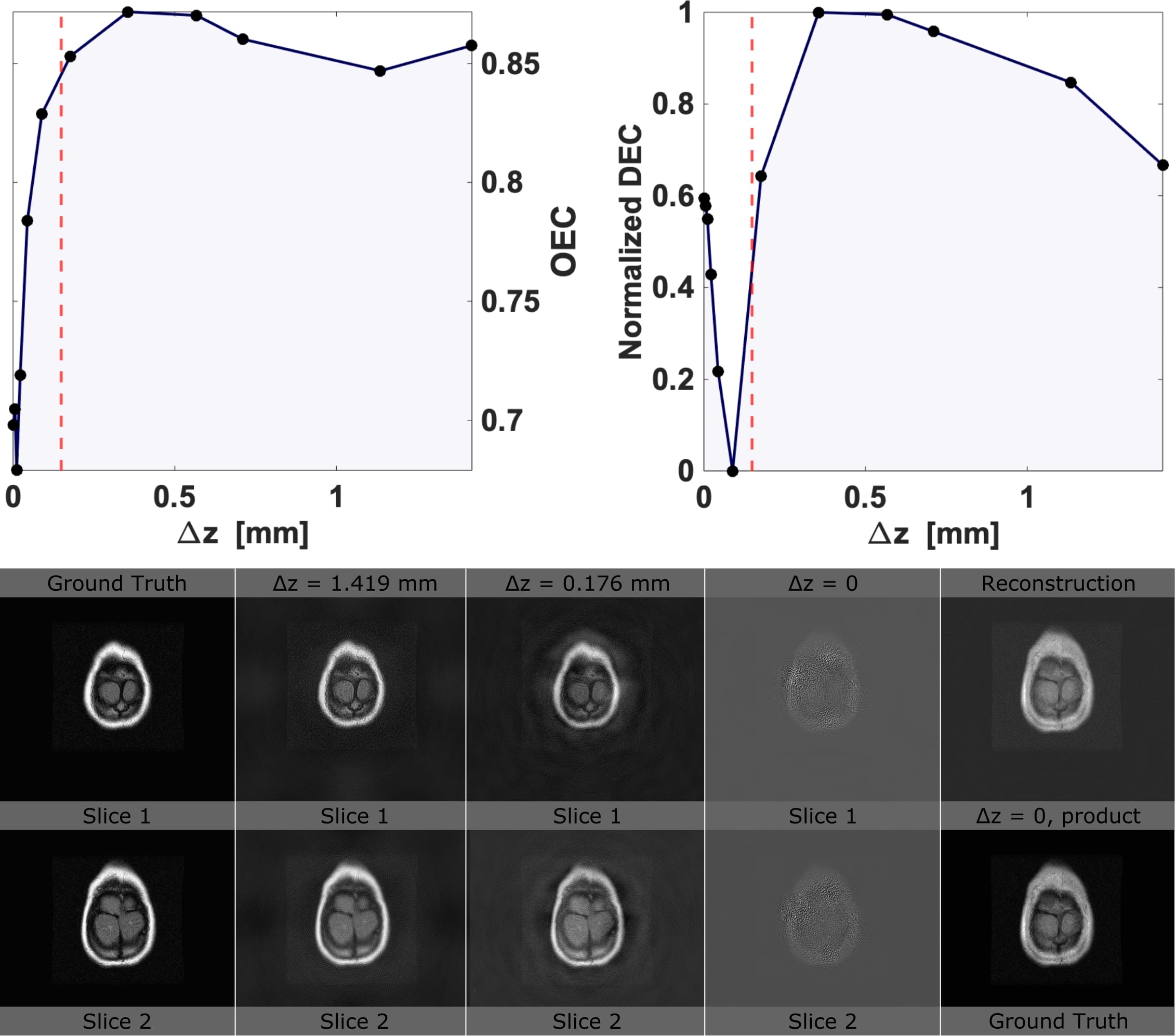}
\caption{Results of the simulated experiments to determine the axial resolution of the 3DSPP system. The final OEC as a function of the separation between the two slices is shown in the top left graph, while the final normalized DEC is shown in the top right. These error metrics are calculated as described in the text and the dashed red line in each represents the calculated axial resolution, 150 $\mu$m. The images below show the slices used to simulate ideal diffraction patterns on the left, which we call the ground truth, and the resultant reconstructed slices at the specified axial separations. These results were selected because they exemplify the general behavior of the reconstructions at 0 separation, just above and well above the axial resolution. The bottom right most panels show the product of the recosntructed slices for $\Delta$z = 0 (above), and the product of the ground truth slices (below).}
\label{fig:axialResSim}
\end{figure}

To test the axial resolution in Eqn. \ref{eq:axialRes}, we performed a series of simulated experiments. We simulated two slices of another example MRI data set from MATLAB and varied the separation between the slices. To quantify the accuracy of the reconstructions we developed two error metrics. The object error complement (OEC) is calculated as one minus the RMS difference between the absolute value of each pixel in each slice of the simulated and reconstructed object. Similarly, the diffraction error complement (DEC) is calculated as one minus the RMS difference between the absolute value of the intensity pattern produced by the simulated and reconstructed slices. The results of these simulations are shown in Fig. \ref{fig:axialResSim}. They show that our calculated axial resolution is accurate because both error metrics exhibit dramatically different behavior above and below 150 $\mu$m. 
 
 The OEC is effectively constant until the separation between the slices reduces below the axial resolution, where it decreases monotonically. The diffraction error compliment data presented in the top right panel of Fig. \ref{fig:axialResSim} is normalized such that 1 is the maximum DEC and 0 is the minimum. Like the OEC, the DEC decreases dramatically as the slice separation decreases below then axial resolution limit. However, as that separation continues to decrease the diffraction error compliment increases again. This unexpected result can be explained by considering that once the separation between the slices is so small that the propagation of light between them becomes negligible, the diffraction pattern from two separated slices is effectively the same as the diffraction pattern from one slice comprised of the product of the two individual slices. The reconstruction of two slices with no separation ($\Delta$z = 0) is shown in Fig. \ref{fig:axialResSim}. There are no discernible details of either slice in the reconstructions, but the product of the reconstructed slices matches the product of the ground truth slices. This corroborates the above explanation of the behavior of the DEC below the axial resolution limit, and justifies the claim that the microscope is diffraction limited in the axial direction. Moreover, the simulated axial resolution matches the theoretical value, which gives confidence in Eqn. \ref{eq:axialRes}. There are a few parameters in Eqn. \ref{eq:axialRes} that can be optimized for better resolution in the single-shot system. For instance, we could theoretically push the resolution of the system to less than 10 $\mu$m by using lenses in the 4f imaging system with half the focal length and doubling the size of the detector, which would also double the size of the diffraction patterns. Making these changes would give axial resolution 16 times smaller than the current system achieves. Axial resolution on the order of 10$\mu$m is ideal for many applications, including plasma imaging.

\begin{figure*}[htbp!]
\centering
\includegraphics[width=\linewidth]{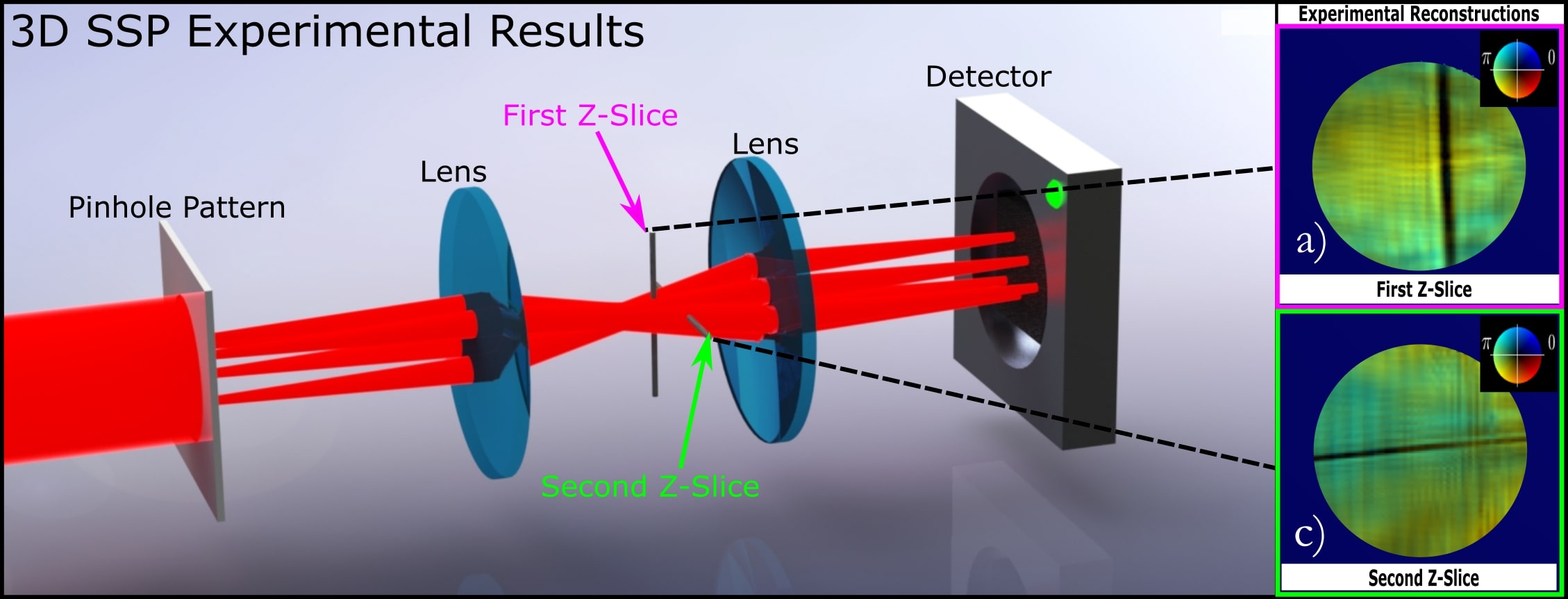}
\caption{setup and results from experimentally verifying the 3DSS technique. On the left, a diagram shows the experimental setup of the 3DSSP system. The lenses (5cm focal length) make a 4-f imaging system that images the DOE (pinhole pattern) to the detector. The object is comprised of 2 hairs that are orthogonally oriented and separated by 5mm. The first hair was placed 5mm out of the focus of the first lens.  Images of the reconstructions of the two slices are shown on the right. The top image shows the amplitude (brightness) and phase (color) of the first z-slice, while the bottom image shows the amplitude and phase of the second z-slice. The algorithm was able to deconvolve the two slices such that the first slice only shows a vertical hair and the second slice only shows the horizontal hair.}
\label{fig:expResults}
\end{figure*}

\subsection{Experimental Results}
We experimentally verified 3DSSP using our existing SSP system. The DOE is a Fermat spiral of 40 circular pinholes of 55 $\mu$m diameter which is illuminated by a large isotropic 532 nm collimated beam. The 4f imaging system has unit magnification and $f$ = 5 cm lenses, and the detector is a Thorlabs 8051M-USB 8 megapixel monochrome CCD camera with the face-plate and wedge window removed. The detector has 3296x2472 square pixels with a $dX$ = 5.5 $\mu$m pitch. The DOE slightly under-filled the detector such that it only has a square effective area of about 2000 pixels. The experimental setup is shown on the left of Fig. \ref{fig:expResults}.

To create an object for testing 3DSSP, we used two orthogonally oriented ~50 $\mu$m thick strands of hair axially separated by 5 mm in a cross-like pattern. The discrete two plane nature of the object was chosen because it makes identification of a successful reconstruction trivial. We determined the crossover point and mounted the object on a micrometer such that the first hair was at the crossover plane. Then we used the micrometer to move the object such that the first hair was 5 cm downstream of the crossover point, $\delta$ = 5 cm, where the overlap fraction calculated by Eqn. \ref{eq:axialDist} is $\eta$ = 0.66.

Diffraction patterns were collected using multiple exposure times that were stitched together using a high dynamic range (HDR) algorithm, which improved the fidelity of the reconstructions for this proof of principle. After collecting the diffraction pattern data, we applied the centroidal Voronoi tessellation to break up the detector into 40 segments of ~220x220 pixels. We calculated the initial positions of the beamlets on the first object slice. These were determined by removing the object, imaging the DOE directly to determine beamlet positions on the detector, and then calculating the transverse shifts for each beamlet as described above. 

The diffraction patterns and calculated positions on the first slice were fed into the 3DSSP algorithm and the reconstructions are shown on the right of Fig. \ref{fig:expResults}. As shown in the figure, the algorithm successfully deconvolves the first slice which shows the vertical bar in the figure form the second slice which shows a horizontal bar. The modulations in the reconstruction are a result of imperfect knowledge of the distance between strands, $\delta$, and a limited number of iterations. The reconstruction of this relatively simple object demonstrates that experimental 3DSSP can successfully deconvolve multiple 2D slices of a 3D object. Future work will expand on this experimental proof of concept by increasing the number of reconstructed slices, determining the resolution of the system experimentally and establishing the minimum transmission and/or maximum extent of an object that can be imaged. 

\section{Conclusion}
By exploiting a feature of the SSP setup, that the position of the beamlets change as a function of distance from the crossover point, we developed a novel algorithm for three-dimensional single-shot ptychographic imaging. While 3D ptychography techniques exist, the geometry of single-shot ptychography setups lends itself well to create highly constrained systems for reconstructing objects at multiple z-slices, and allows for time resolved imaging. Through a number of simulations, we showed the power of the 3DSSP algorithm to image 3D objects with high reconstruction quality. The axial resolution of this technique was derived and found to be consistent with the depth of field of a standard microscope. Simulations support the axial resolution of 150 $\mu$m and a reconstruction volume of ~ 75 mm$^3$given our setup parameters. The axial resolution can be improved down to around 10 $\mu$m with reasonable modifications to the setup. The 3DSSP technique was experimentally realized by imaging orthogonally oriented hairs separated by 5 mm. The reconstructions of each hair were deconvolved as predicted by our simulations. Through both simulated and experimental reconstructions, we have shown that 3DSSP is a reliable method for imaging a multitude of 3D objects. This is currently the only method to image transient, non-reproducible objects in three-dimensions in both phase and amplitude and with temporal resolution set by the illumination. Metrologies such as this will be imperative for studying phenomena such as dynamically evolving plasma.

\section*{Funding Information}
The authors gratefully acknowledge funding from the Air Force through AFOSR FA9550-18-1-0089 and Los Alamos National Laboratory through contract number 501188.

\section*{Acknowledgments}

The authors thank Professor Jeff Squier for fruitful discussions regarding the development of three-dimensional single-shot ptychography.

\section*{Disclosures}

\noindent\textbf{Disclosures.} The authors declare no conflicts of interest.



\section{Figures}

\end{document}